\documentclass[10pt,conference]{IEEEtran} 
\IEEEoverridecommandlockouts
\usepackage{cite}
\usepackage{tikz}
\usepackage{multirow}
\usepackage{hyperref}
\usepackage{amsmath,amssymb,amsfonts}
\usepackage{algorithmic}
\usepackage{graphicx}
\usepackage{textcomp}
\usepackage{xcolor}
\usepackage{xspace}
\usepackage{comment}
\usepackage{todonotes}
\def\BibTeX{{\rm B\kern-.05em{\sc i\kern-.025em b}\kern-.08em
    T\kern-.1667em\lower.7ex\hbox{E}\kern-.125emX}}
\def\FS{${\mathcal F{\kern -0.2em S}}$\xspace}
\def\CoT{${\mathcal C{\kern-0.15em o}{\kern-0.15em T}}$\xspace}
\def\SC{${\mathcal S{\kern-0.10em {\text -}}{\kern-0.10em C}}$\xspace}

\definecolor{Gray}{gray}{0.3}
\tikzstyle{mybox} = [draw=black, very thick, rectangle, rounded corners, inner ysep=5pt, inner xsep=5pt, fill=gray!20]

\newcommand{\xyz}[2]{
    \smallskip
    \noindent
    \begin{tikzpicture}
        \node [mybox] (box){%
        \centering
        \begin{minipage}{.465\textwidth}
        \fontsize{8.8}{10}\selectfont
        \textbf{RQ #1}. #2
        \end{minipage}
        };
    \end{tikzpicture}%
}

\begin{document}

\title{Better patching using LLM prompting, \emph{via} Self-Consistency
\thanks{This material is based upon work supported by the National Science Foundation under Grant NSF CCF (SHF-MEDIUM) No. 2107592. Any opinions, findings, and conclusions or recommendations expressed in this material are those of the author(s) and do not necessarily reflect the views of the National Science Foundation.}
}


\author{
\IEEEauthorblockN{Toufique Ahmed}
\IEEEauthorblockA{
% \textit{University of California, Davis}\\
\textit{UC Davis,  California, USA}\\
%Davis, USA \\
tfahmed@ucdavis.edu
}
\and
\IEEEauthorblockN{Premkumar Devanbu}
\IEEEauthorblockA{
%\textit{University of California, Davis}\\
\textit{UC Davis,  California, USA}\\
%Davis, USA \\
ptdevanbu@ucdavis.edu
}

}

\maketitle

\begin{abstract}
Large Language models (LLMs) can be induced to solve non-trivial problems
with ``few-shot" prompts including illustrative \emph{problem-solution} examples. 
Now if the few-shots also include ``chain of thought" (\CoT) explanations, which
are of the form \emph{problem-explanation-solution}, LLMs will 
generate a ``explained" solution, and perform even better. Recently an exciting, substantially better technique, self-consistency~\cite{wang2022self} (\SC) has emerged, based on the intuition that there
are many plausible explanations for the \underline{right} solution; when the LLM is sampled
repeatedly to generate a pool of explanation-solution pairs, for a given problem, the most frequently occurring \emph{solutions} in the pool (ignoring the \emph{explanations}) tend to be even more likely to be correct! 

Unfortunately, the use of this highly-performant \SC (or even \CoT)  approach in software engineering settings is hampered by the \emph{lack of explanations}; most software datasets lack explanations. In this paper, we describe an application of the \SC approach to program repair, using the \emph{commit log} on the fix as the explanation, \emph{only} in the illustrative few-shots. We achieve state-of-the art results, beating previous approaches to  prompting-based program repair, on the MODIT dataset; we also find evidence suggesting that the correct
commit messages are helping the LLM learn to produce better patches. %\todo{really SOTA?} {In Prompting base I think we are achieveing SOTA}

\end{abstract}


\section{Introduction}
For more than a decade, language models have found many applications
in the field of software engineering. They are based on a simple idea:
given a context (or a $prompt$), try to predict the next token (or a
missing one); in other words, learn a conditional probability distribution
of the form $p(token \mid prompt)$. In neural models, the prompt is a sequence
of tokens that is internally represented by a high-dimensional vector,
which encodes the parameters of the neural computations. By repeatedly
applying this conditional distribution, we can generate sequences of tokens
that depend on the prompt and the previously generated tokens (\emph{aka}
autoregressive generation).

%TThese models have proven remarkably adaptable to varied tasks. 
Modern, instruction-tuned neural language models such as GPT-3 and LLAMA (colloquially known as ``LLMs"), can have hundreds of billions of parameters; so when representing a $prompt$ internally, they do so in a very rich and complex space, and then undertake conditional auto-regressive generation starting thereon. The richness of the space for representing prompts allows a wide range of prompt construction, leading to an entirely new field of ``prompt engineering"; by guiding LLMs to specific regions of the ``context-space" prior, different auto-regressive generation possibilities,  conditioned on this prior could ensue, leading to different ways of solving the actual task that one encodes in a prior.
Approaches include few-shot learning~\cite{brown2020language}, and chain-of-thought~\cite{wei2022chain}.  
``Few shot" (\FS) learning amounts to providing examples (input/output \emph{pairs}) illustrating the task within the prompt, and then asking for the output for a target input; 
``chain-of-thought" (\CoT) amounts to providing \emph{reasons} connecting input and output pairs, thus input-reason-output \emph{triples}. 
\FS and \CoT are complementary. All these techniques work in software engineering tasks.   

 Since these generative possibilities amount
to sampling sequentially from the next-token distribution that the LLM has learned, different sequences could be generated. Normally, one follows a \emph{greedy sampling} approach, taking the most likely token at each stage of the auto-regressive generation. % one could also take a beam-search approach, and choose a most-likely sequence out of a generated possibility pool of fixed size. 

In a recent paper~\cite{wang2022self} Wang \emph{et al} proposed a different approach, to selecting an output sequence from an auto-regressively trained model, 
called ``self-consistency" (\SC). With \SC, the idea is that the model is prompted (using \CoT, perhaps with
%\ifthenelse{}
\FS) to produce \emph{first} an \emph{explanation}, and then an \emph{answer} $\ldots$ but for \SC, the 
LLM  is sampled repeatedly, using a ``high temperature"\footnote{Temperature $t, t \in [0,1]$ here refers to the likelihood of choosing a next token that's not necessarily the most likely following token as per the model. $t = 0$ is just greedy selection.}.
This repeated sampling can be thought of as a way to model to generate completions (first \emph{explanations}, then \emph{answers}) from ``different perspectives". From these varied ``perspectives", 
\SC simply chooses the most frequent \underline{{\em answer}} (thus ignoring, or ``marginalizing over'' the explanations). \SC is remarkably effective, showing significant improvements on many tasks in Natural language. In this paper, we explore application of this idea to program repair.

Our primary contributions are: 
\begin{enumerate}
\item We find evidence suggesting that \SC, with commit-log message
as the reason, improves performance on the code-repair task. 
\item Our data suggests that using the correct commit log messages within
the few-shots actually does help produce sigificantly better answers, with \SC; random  commit messages do not.
\end{enumerate}

\begin{figure*}[t]
    \centering
        \caption{Steps followed for \FS,\CoT, and \SC in program repair.}
    \label{pipeline}
    \includegraphics[scale=0.35]{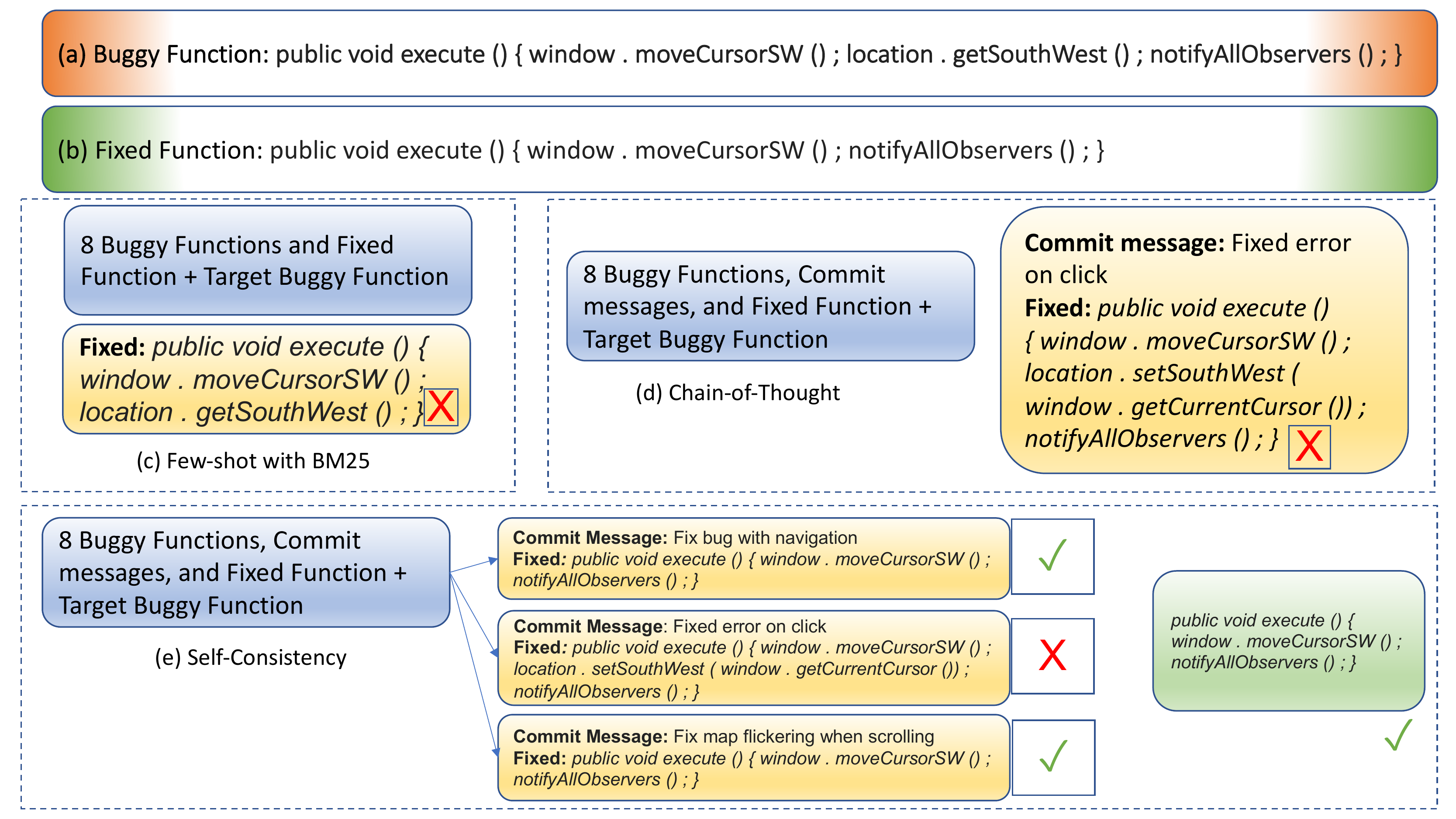}
\vspace{-.5cm}
\end{figure*}

%\vspace{-1cm}

\section{Background \& Related work}
\label{background}
\subsection{Few-shot Learning}
\FS learning is very useful with  (LLMs)~\cite{brown2020language,chen2021evaluating}. With \FS learning, \emph{e.g.,}
% we provide the models with a set of examples and ask them to perform a specific task based on the given input. For instance, 
if we desire to translate English to German, we assemble a few English-input, German-output pairs into  a $prompt$.  Now if a final English sentence is added to the prompt, the model completes the prompt with German translation theoreof. This trick works for a range of tasks; remarkably, the model does this, \emph{without requiring any weight updates}. While \FS learning 
%was initially introduced 
originated in NLP applications, it also works for 
a wide range of software engineering tasks, including code summarization~\cite{ahmed2022few}, code repair~\cite{nashid2023retrieval}, assertion generation~\cite{nashid2023retrieval}, code mutation~\cite{bareiss2022code}, test oracle generation from natural language documentation~\cite{bareiss2022code}, and test case generation~\cite{bareiss2022code}. Notably, using \FS on these tasks
appears to improve performance compared to previous state-of-the-art models. 
%\subsection{\FS{} Learning with BM25}
%\FS{} learning has emerged as a promising approach, particularly for new tasks or tasks with limited samples, as it reduces the dependence on fine-tuning with thousands of samples. 
\FS learning is felicitously helpful when even just a few labeled/curated examples are available; but
for many SE tasks, data is often mined from repositories, and is more abundant. Can this more
abundant data be leveraged in a few \FS setting? 
%However, what about scenarios where we already possess a substantial amount of data for fine-tuning? Should we disregard these additional samples or find a way to incorporate them to achieve better performance? It is important to note that due to the limitations of model length, we can only utilize a few samples for each task.

Nashid et al.~\cite{nashid2023retrieval} used %conducted a study demonstrating that if we employ 
an IR-based approach (BM25~\cite{robertson2009probabilistic}) to select specific few-shot samples, which improved \FS performance for both code repair and assertion generation. Ahmed et al. found that this approach also improves  code summarization performance~\cite{ahmed2023improving}. 
%BM25 is a frequency-based relevance ranking algorithm, which is a variant of TF-IDF~\cite{ramos2003using}. The researchers experimented with various retrieval approaches but found BM25 to be the most effective one.  
Fig.~\ref{pipeline}-(c) illustrates the input-output of the BM25-based few-shot learning method for program repair task, where eight (can be more or less depending on the context length) pairs of buggy-fixed function examples were retrieved using the BM25 algorithm and concatenated with the buggy version of the target function. Consequently, the fixed version of the function was retrieved from the model's output.

\subsection{Chain-of-Thought (\CoT)}
Wei et al.~\cite{wei2022chain} found that %conducted an investigation on how the 
forcing the generation of intermediate reasoning steps (\emph{viz.,} Chain of Thought (\CoT)), %significantly enhances the capabilities of large language models 
substantially improves LLM performance on
%in performing 
complex reasoning tasks. As with \FS learning, for \CoT,  we prompt the model with a few input-output pair ``shots"; but now, each shot is augmented with an intervening \emph{reasoning path}: each shot is now an ordered triplet  of $\langle input, reasoning{\text -}path, output \rangle$. 
A prompt includes several triplets, followed by a target $input$; the model then generates the $reasoning{\text -}path$, and the $output$. This approach substantially improves task performance
in several settings, including
%The generation of this reasoning (Chain of Thought) enables the model to achieve improved problem-solving abilities. It is worth noting that this approach has been successfully applied to tasks involving 
commonsense, and symbolic reasoning. %, even without relying on a retrieval-based \FS approach like BM25.
%However, its implications in the Software Engineering (SE) domain remain unexplored.

But this approach requires atleast \emph{some} few-shots with $reasoning{\text -}path$s.
In general, SE datasets (for repair, summarization, code-retrieval \emph{etc}) have
abundant $\langle input, output \rangle$ pairs, but  $reasoning{\text -}path$s are rare. 
Creating these paths, for few-shotting, after-the-fact, can be a challenge. 
%In the context of program repair, generating or writing the exact reason for each test sample can be challenging. 

We posit that \emph{a summary or commit log message could serve as a reasoning path}. 
%However, an approximate representation of the reason behind the change can be achieved by utilizing commit messages. 
Fig.~\ref{pipeline}-(d) illustrates the model output when the model is prompted to generate the reason/commit message prior to generating the fixed function. The MODIT~\cite{chakraborty2021multi} work used commit messages in program repair tasks. However, in MODIT, commit messages \emph{associated with the {\bf target} $input$} were actually inserted in the prompt\footnote{The MODIT approach arguably is not leaking test-data, since it's very plausible that a
developer fixing a bug knows what needs to happen, and can thus write commit log, before they actually code-up the change.}. 
%were used during the test time. 
As we shall see next, for the \SC approach, the reasoning path (commit message) associated
with the target \emph{must also be}
generated, in order to allow model some additional randomness to subsequently generate
a range of different possible outputs for the target input. 

\begin{table*}[t]
\centering
\caption{Performance of Chain-of-thought and Self-consistency in program repair task}
\label{rq1}
\resizebox{.85\textwidth}{!}{%
\renewcommand{\arraystretch}{1.2}% Tighter
\begin{tabular}{lcccccccccc}
\hline
\multicolumn{1}{c}{Dataset} & Greedy & \CoT & \SC     & \begin{tabular}[c]{@{}c@{}}Relative Gain over \\ Greedy\end{tabular} & p-value & Greedy + BM25 & \CoT+ BM25 & \SC+ BM25      & \begin{tabular}[c]{@{}c@{}}Relative Gain over \\ Greedy + BM25\end{tabular} & p-value \\ \hline
$B2F_s$                & 9.50\% & 10.00\%  & 13.50\% & +42.10\%                                                    & $<0.01$      & 29.00\%     & 29.00\%  & 31.80\% & +9.65\%                                                          & $<0.01$     \\
$B2F_m$               & 11.20\%     & 10.40\%  & 15.50\%      & +38.39\%                                                          & $<0.01$      & 19.10\%     & 20.20\%  & 21.60\% & +13.08\%                                                         & $<0.01$     \\ \hline 
\end{tabular}
}
\vspace{-.4cm}
\end{table*}

\subsection{Self-Consistency (\SC)}

Wang et al.~\cite{wang2022self} introduced self-consistency-based generation, which improves over the naive greedy decoding approach used in chain-of-thought prompting. \SC
starts with a prompt with few-shot \CoT triples ending with a target $input$, thus prompting the generation of a $reasoning{\text -}path$ and an $output$. So far, similar to conventional \CoT. 

But now, \SC posits higher-temperature generation, thus sampling a \emph{collection} of distinct reasoning paths and outputs, rather than greedily sampling (with `Temperature 0') a single reasoning path and output. From this collection, they select the most frequently occurring output (ignoring the reasoning paths). \SC builds upon the intuition that the same correct answer to a complex problem is often reached \emph{via} several distinct reasoning approaches. The \SC expansion of  \CoT prompting is remarkably effective. In this study, we generate upto  50 samples for each input sample, and the final model output is determined through majority voting. Fig.~\ref{pipeline}-(e) showcases an example of self-consistency with three samples.

It's important to note here that \SC requires the generation of a 
reasoning-path \emph{after} the target input in the prompt, and \emph{before} 
the required output. To prompt an LLM to do this, we do require the same
kind of triplet as used in \CoT. Unfortunately, in SE, while datasets for  tasks include inputs and outputs, $reasoning{\text -}path$ element is almost
never available. Our idea here was to use commit-log messages as
the $reasoning{\text -}path$, and then apply \SC to see if 
new SOTA performance can be achieved for program-repair, on the MODIT dataset. 
%\todo{Need to introduce RQs better}

\subsection{Research Questions (RQs)}
Our first question considers the value of the \SC approach for program repair tasks.  
We consider the setting where the model's input is the complete buggy function \emph{without} bug localization, and the output is the complete fixed function. Currently, 
BM25-based few-shot retrieval has reached a high-water mark~\cite{nashid2023retrieval} in this setting. We study if BM25, when combined  with self-consistency, delivers even better performance. 
\xyz{1}{Does \SC improve program repair performance?}
When using \SC, one has to sample a number of reasoning paths, over which to find
consistent outputs. More paths might lead to better consistency; but computational costs scale also linearly with more paths, so it
would be good find the balance. 
%In this research, our objective is to determine the optimal number of samples required to effectively apply \SC. We aim to strike a balance between the effectiveness of \SC and the associated costs and time requirements involved in generating a higher number of samples.
\xyz{2}{How does performance vary with the number of generated reasoning paths?}
Finally: our few-shot examples include the commit message, actually
associated with the bug-fix commit, as the ``reasoning path", to push the model to generate a good reasoning-path for the target input.  But are these \emph{actual} commit messages
matter? Is the model actually learning from these reasoning paths in the few-shots?  To study this question, we prepared few-shot samples where the commit message was \emph{not} the one associated with the fix, but \emph{some other random} commit message sourced from the data. 
If commit messages are indeed useful ``reasoning paths", such random ``reasoning paths" 
\emph{should} confuse the model and diminish performance. 
\xyz{3}{How does the performance change with use of random commit messages instead of the original ones?}
%\begin{enumerate}
%    \item \textbf{RQ1:} Does \SC improve program repair performance?
%    \item \textbf{RQ2:} How does performance vary with the number of generated reasoning paths?
    %\item \textbf{RQ3:} $\ldots$ vary with the number of samples in BM25 pool?
%    \item \textbf{RQ3:} How does the performance change with use of random commit messages instead of the original ones?
%\end{enumerate}

\section{Methodology}
%We now provide a brief overview of the dataset, and our proposed approach.

\subsection{Dataset}

We use the MODIT dataset, which includes two subsets ($B2F_s$, which has smaller sequences, and $B2F_m$)~\cite{tufano2019empirical}. MODIT comprises bug-fix commits (including \emph{commit-logs}), from GitHub. Variants of these datasets have been used in evaluating NatGen~\cite{chakraborty2022natgen}, CodeT5~\cite{wang2021codet5}, and also in the CodeXGLUE~\cite{lu1codexglue} benchmarks. Other work~\cite{nashid2023retrieval} used the  TFix~\cite{berabi2021tfix} dataset, which is classified into the 52 error types flagged by ESLint~\footnote{See \url{https://palantir.github.io/tslint/}}. The MODIT dataset, includes commit messages written by developers;
we use these commit-messages as the Chain-of-thought (which is required
if one seeks to apply \SC to a task). 
%and facilitates the application of \CoT and \SC to the program repair problem. 
In our experiment, we use the training partition as the sample pool from which BM25 retrieves relevant few-shot samples. %\todo{{\bf Toufique} check this} \toufique{seems okay}
From the test partition, we randomly sampled 1000 examples for our tests, to gain
statistical power while also dealing with OpenAI's rate limitations \& associated costs. 

Program repair is a popular topic; conducting a comprehensive comparison of all the models~\cite{chen2019sequencer,jiang2021cure,lutellier2020coconut,ye2022neural,jiang2023impact} and datasets~\cite{berabi2021tfix,just2014defects4j,lin2017quixbugs} is beyond the scope of this paper; we're interested to see if commit log messages can
be used as chain-of-thought for applying self-consistency. 

\subsection{Model}
One can access models of varying sizes, that are trained using code and related NL descriptions. Models like Codex~\cite{chen2021evaluating}, PaML~\cite{chowdhery2022palm} PolyCoder~\cite{xu2022systematic}, and CodeGen~\cite{nijkamp2022codegen} have gained significant popularity. Our primary focus is on the Code-DaVinci-002 model, which has 175 billion parameters and demonstrate exceptional performance in code-related tasks. %Additionally, we employ the Code-Cushman-001 model (with 12 billion parameters) to study  scalability. 
We access this model \emph{via} OpenAI API.

\subsection{Proposed Approach, Baseline \& evaluation criteria}

We base-line \SC, on the program repair task, over the state-of-the-art Nashid \emph{et al}~\cite{nashid2023retrieval}, which uses BM25-retrieved few-shots, with
greedy decoding. 
We tried several \FS decoding techniques to generate correct patches:
greedy decoding (temperature 0), \CoT, and \SC. Greedy-decoding just sequentially selects the most probable output. \CoT also decodes greedily:  but it uses triplet-shots 
in the few-shot prompt to induce the model to follow-up the target input with a reasoning-path before generating the output. For \SC, we use a temperature of 0.7 to configure the model to generate more diverse reasoning-paths \& outputs for the target input. 
We sampled up to 50 sequences for each of the 1000 target test cases during the experiment. However, when reporting the results, we focused on the first 30 samples: we found that the performance tends to plateau after the initial 10 samples (Section~\ref{samplecount}). As with~\cite{nashid2023retrieval}, BM25 was used to find relevant few-shots. We repeated all three approaches (greedy, \CoT and \SC), %\todo{This is a bit confusing}
%but instead of using the same 8 samples, 
using samples chosen by BM25. Using the %\todo{CHECK!! Toufique: We can call it SOTA} 
SOTA approach~\cite{nashid2023retrieval}
(BM25) as a baseline, we specifically examine the benefits of using \SC in conjunction with BM25. We used the exact match as evaluation metric.

\begin{figure}[t]
    \centering
    \caption{Number of \SC samples vs. Acuuracy}
    \label{samples}
    \includegraphics[scale=0.25]{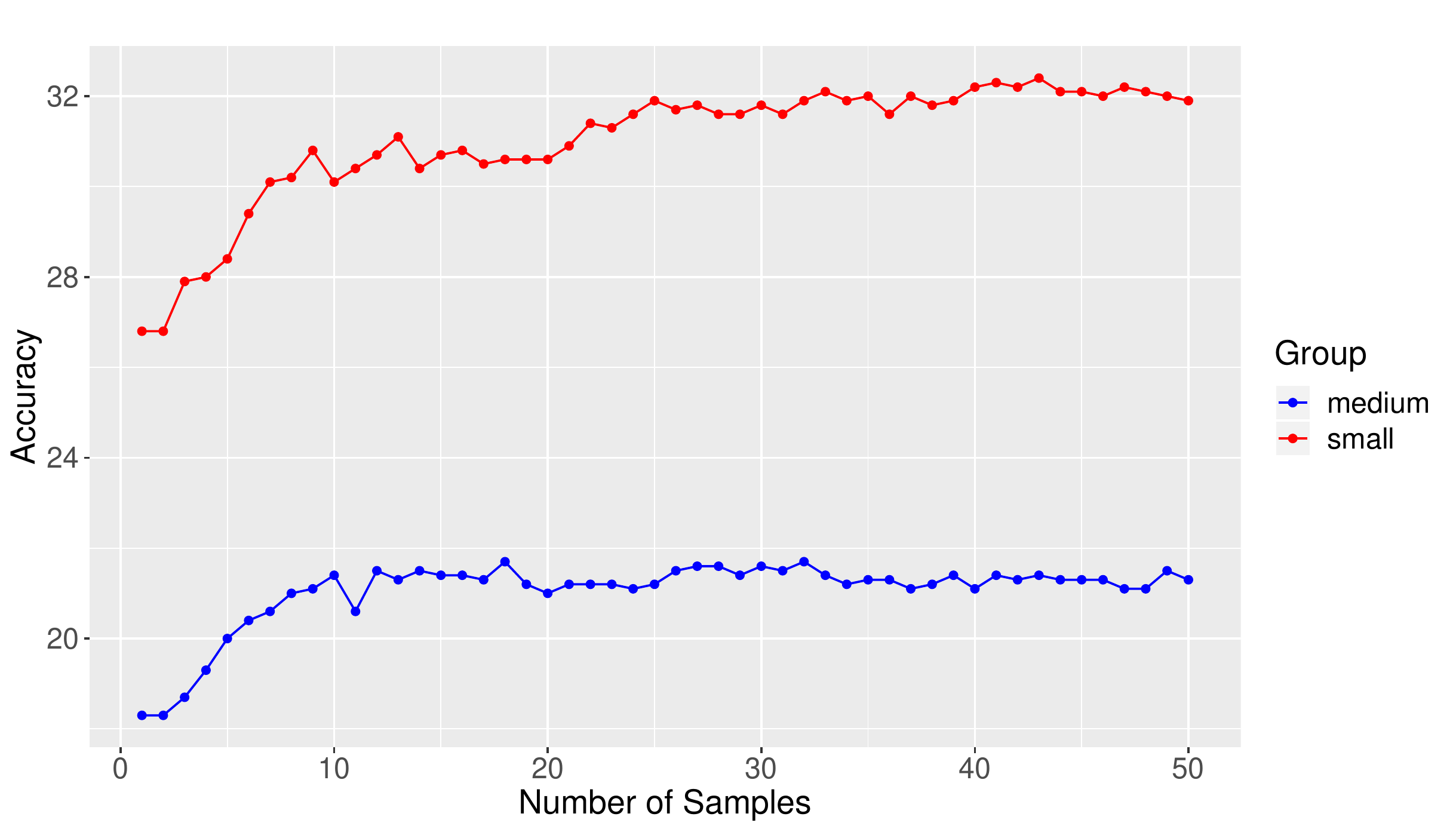}

\end{figure}

\section{Result}

\subsection{\textbf{RQ1:} Performance of the proposed approached}
Table~\ref{rq1} shows the performance (top-1 exact match with the fix)  of \CoT and \SC in the program repair task.  Using BM25 to select few-shots works best.  We also show results for the traditional fixed 8-sample \FS learning: \SC improves over greedy decoding (\emph{sans} BM25) by 42.10\% and 38.39\% (relatively). However, when few-shots are selected using BM25, the gains are lower, 9.65\% and 13.08\% respectively. To determine statistical significance, we performed a McNemar test~\cite{mcnemar1947note} (a non-parametric test used to analyze matched nominal data). For each pair of settings above,  self-consistency improves ($p < 0.01$) over greedy decoding. We also observe that the using \CoT sometimes provides improvement, albeit not sufficient to be statistically significant. 
%  improvements \todo{Statistically significant
%in this case? toufique: no statistical significance}

\subsection{\textbf{RQ2:}How many samples are needed for \SC?}
\label{samplecount}

%The \SC mechanism relies on generating multiple solutions from the model, which can be a costly and time-consuming process; s
To manage costs of repeated sampling, we study how performance changes with number of samples. Fig.~\ref{samples} suggests that the performance consistently improves for the first 10 samples for both datasets. However, for $B2F_m$, we saw no significant benefits beyond 10 samples. On the other hand, for $B2F_s$, the performance improves with an increasing number of samples, though the benefits taper off.  To ensure generality in our reporting, we consider 30 samples as the chosen number for our results and the complete processing takes less than 5 seconds for each sample (ignoring rate limitations). %\todo{Need to provide

\subsection{\textbf{RQ3:} Random commit message vs. original ones}

\begin{table}[t]
\centering
\caption{Impact of using random commit messages.}
\label{random}
\resizebox{.75\columnwidth}{!}{%
\renewcommand{\arraystretch}{1.2}% Tighter
\begin{tabular}{lccc}
\hline
\multicolumn{1}{c}{\multirow{2}{*}{Dataset}} & \multirow{2}{*}{Acc.} & \multicolumn{2}{c}{Comparing with baseline} \\
\multicolumn{1}{c}{}                         &                       & Gain               & p-value            \\ \hline
$B2F_s$                                         & 30.00\%                & +3.44\%             & 0.30   \emph{(not Sig.)}           \\
$B2F_m$                                         & 19.90\%                & +4.18\%             & 0.30   \emph{(not Sig.)} \\ \hline            
\end{tabular}
}
\vspace{-.4cm}
\end{table}

We tried pairing the buggy program with random commit messages instead of the commit messages in \FS samples. We saw lower improvement of 3.44\% and 4.18\%  over our baseline approach; moreover, statistical significance was not achieved (Table~\ref{random}). %Furthermore, when comparing it with the original commit messages, we observed performance drops of 5.66\% and 7.87\%. 
%Thus, using random commit messages improved slightly over the baseline vanilla BM25 model, but not enough to be statistically significant. 
However, using the original or correct commit messages resulted in statistically significant improvement: this suggests
that the LLM is learning better reasoning paths from the original commit
messages, for the task of program repair.

\section{Discussion}
%\todo{This is focused on Natgen. we need to include Codit5 and other other models
%in tis discussion. e.g.:  "We note that our approach is based on the few-shot %prompting of LLMs, without the fine-tuning used in Natgen, Codit5 etc. Fine-%tuning
%actually changes the parameters of the language model to adjust the model to
%be better tuned on-task training sets; few-shotting doesn't change the %parameters. Fine-tuning massive LMs like Davinci requires resources beyond %the reach of all but a  few select institutions.  
%Nevertheless, our few-shot approach beats some earlier fine-tuned approaches %like Natgen, but doesn't beat others like Codit5. Future work could determine %whether fine-tuning
%a Codex-scale model can beat earlier fine-tuned models}
%\todo{will discuss about the bert and other model and some threats}
Before LLMs, encoder-decoder models were commonly employed to fix buggy programs. These models were fine-tuned using the entire training data and exhibited satisfactory performance on this specific task. A recent model called NatGen~\cite{chakraborty2022natgen} scored an accuracy of 23.43\% and 14.93\% on $B2F_s$ and $B2F_m$, respectively. These numbers are lower than our scores of 31.80\% and 21.60\% accuracy; note that Natgen \emph{included} the commit log 
message for the test example, we don't; we \emph{generate} it, but then discard it, and retain just the fix code.  It is important to note that our performance 
wasn't measured on the full test set, rather on just a sample; 
%cannot be directly compared to NatGen or prior encoder-decoder models since we utilized a subset of the test data. 
Nevertheless, since
we randomly chose a  large (1,000) sample for our experiments, one can 
reasonably expect that our findings are robust (note $p < 0.01$ as
per McNemar test for our main finding). 

%\footnote{The binomial 
%one-sided lower-bound
%with 1000 trials, at the 99\% confidence level is 28.4\% and 18.6\% respectively,
%higher than 23.4\% and 14.9\%}. 

The commit messages in MODIT are sometimes uninformative, and %are often 
limited to ``Bug Fixed''. 
With commit messages of better quality, the model's performance with \CoT and \SC could potentially improve. It is possible to generate informative commit messages with the assistance of other models, but further research is required. 
%However, we acknowledge that exploring this aspect is beyond the scope of our current study, and we leave it for future research.
Several studies have reported higher performance compared to our approach~\cite{chakraborty2021multi,zhang2022coditt5}. However, we note that these works assume that the location of the bug is known, and they generate fix ``snippets" instead of complete fixed functions. This assumption may limit use if the fault location is not known;  
we, however, pass in the whole buggy function to the model~\cite{chakraborty2021multi}, without fault localization, and
require the entire fixed function as output.

%\section{Related Work}

\section{Conclusion}
This paper explores the idea of using self-consistency for defect repair,
specifically using the commit-log message as the ``reasoning path" just within
the few-shot illustrative examples (but no commit-log message %\emph{not}
for the test example). We find that this approach provides improved 
performance; furthermore, we find that generating a sample-pool of around 30 explanation-answer pairs, and choosing the most frequent answer, works well. 
We also find that evidence suggesting that the LLM is learning
from the \emph{actual, correct} commit-log message in the few-shots: using random commit-log messages doesn't provide any significant improvement over prior approaches. 
For our 
data and Script, please access \url{https://doi.org/10.5281/zenodo.7968641}.

%\section*{Acknowledgment}

%The preferred spelling of the word ``acknowledgment'' in %America is without 
%an ``e'' after the ``g''. Avoid the stilted expression ``one of us (R. B. 
%G.) thanks $\ldots$''. Instead, try ``R. B. G. thanks$\ldots$''. Put sponsor 
%acknowledgments in the unnumbered footnote on the first page.

%This is codex~\cite{chen2021evaluating}
%\newpage

\bibliographystyle{IEEEtran}
\bibliography{ref.bib}

\end{document}